\documentclass[12pt]{article}
\def\e{{\rm e}}
\def\a{\alpha}
\def\Erf{{\rm Erf}}
\usepackage[english]{babel}
\usepackage{amssymb}
\usepackage{amsmath}
\usepackage{indentfirst}
\usepackage{graphicx}
\usepackage{graphics}
\usepackage{epsfig}
\usepackage{subfigure}
\usepackage{multirow}
\usepackage{color}

\newcommand{\mathsym}[1]{{}}

\def\e{{\mathrm{e}}}

\def\alphaa{a}
\def\betab{b}

\def\d{{\rm d}}
\def\e{{\rm e}}
\def\a{\alpha}
\def\Erf{{\rm Erf}}
\def\bfr{{\bf r}}
\def\bfp{{\bf p}}
\def\bfalpha{\boldsymbol{\alpha}}
\def\bfsigma{\boldsymbol{\sigma}}
\def\la{\lambda}
\def\b{\beta}
\def\aa{a}
\def\bb{b}
\def\ala{a}
\begin{document}
\title{QMC approach based on the Bogoliubov independent quark model of the nucleon}
\author{
Henrik Bohr\\
{\it \small Department of Physics, B.307, Danish Technical
University,}\\{ \it \small DK-2800 Lyngby, Denmark}\\Steven A.\ Moszkowski\\
{\sl \small UCLA, Los Angeles, CA 90095, USA} \\Prafulla K.
Panda\\{\it\small Department of Physics, Utkal University,
Bhubaneswar, 751004, India
}
\\Constan\c{c}a Provid\^encia,
Jo\~ao da Provid\^encia\\
{\it \small CFisUC, Departamento de F\'\i sica, Universidade de Coimbra,}\\
{\it \small  P-3004-516 Coimbra, Portugal} }
\date{}
\maketitle

\begin{abstract}
The quark-meson coupling model due to Guichon is  formulated on the
basis of the independent quark model of the nucleon proposed by
Bogoliubov and is applied to the phenomenological descriptions of
symmetric and asymmetric nuclear matter. For symmetric matter, the
model predicts, at saturation density, the incompressibility
$K=335.17$ MeV, the quark effective mass $m_q^*=238.5$ MeV, and the
effective nucleon mass $M^*=
0.76 M,$ where $M$ is the  nucleon mass in vacuum. Neutron star masses
above two solar masses are obtained.
\color{black}


\end{abstract} \section{Introduction}
 About almost half
a century ago,  Bogoliubov proposed an interesting model of baryons
\cite{bogolubov
}, which assumes that these particles are composed of independent
quarks bound by a linearly raising potential, as suggested by gauge
theories. With the help of a single phenomenological parameter, the
string tension $\kappa$, this model is able to qualitatively account
for the dynamically generated mass  of the nucleon, for the
corresponding magnetic moment, and for the mass-radius. The
quark-meson-coupling (QMC) model due to Guichon
\cite{guichon,guichon2} incorporates successfully the quark degrees
of freedom into a many-body effective nuclear Hamiltonian, this
being achieved by describing nucleons in the context of the MIT bag
model \cite{chodos}. The aim of the present note is to obtain a
phenomenological description of hadronic matter in the spirit of the
QMC approach  combined with the Bogoliubov model for the description
of the quark dynamics in the nucleon rest system.
\section{Bogoliubov model}
The independent quark model  of the nucleon proposed by Bogoliubov
\cite{bogolubov} is described by the Hamiltonian
\begin{equation}\label{h_D}h_D=-i\bfalpha\cdot\nabla+\beta\left(\kappa|\bfr|-g^q_\sigma \sigma\right),\end{equation}
where $\beta$ and the components $\alpha_x,\alpha_y,\alpha_z$ of
$\bfalpha$ are Dirac matrices,  $\sigma$ denotes the external scalar
field, $g^q_\sigma$ denotes the quark-$\sigma$ coupling and $\kappa$
denotes the string tension. Contrary to the MIT bag model
\cite{chodos}, in $h_D$
the confining potential for quarks is a Dirac scalar. This fact
motivates and justifies that the QMC approach is revisited in
combination with the Bogoliubov model. The eigenvalues of $h_D$ are
obtained by a scale transformation from the eigenvalues of
\begin{equation*}h_{D0}=-i\bfalpha\cdot\nabla+\beta\left(|\bfr|-\aa\right),\end{equation*}where
$\aa$ is a parameter. The variational principle may be applied, to
the square of $h_D$,
\begin{equation}\label{h_D^2}h_D^2=-\nabla^2+(\kappa|\bfr|-g^q_\sigma
\sigma)^2+i\kappa\beta\bfalpha\cdot{\bfr\over|\bfr|}.\end{equation}
If in $h_D^2$ the term proportional to $\beta$ is neglected, a
simplified model is obtained according to which, and by combining
Bogoliubov \cite{bogolubov} and Guichon \cite{guichon} ideas,
confined quarks in nuclear matter are described, in the nucleon rest
frame, by an equation of the Klein-Gordon type,
$$-\partial^\mu\partial_\mu\psi+(\kappa|\bfr|-g^q_\sigma\sigma)^2\psi=0.$$ The mass squared of the
confined quarks, $m^2$, is the lowest eigenvalue of the simplified
Bogoliubov-Guichon operator
\begin{equation}h_{KG}^2=-\nabla^2+(\kappa|{\bf
r}|-g^q_\sigma\sigma)^2,\label{BKG}\end{equation} and can be easily
expressed in terms  the lowest eigenvalue of the operator
\begin{equation}h_{KG0}^2=-\nabla^2+(|{\bf
r}|-\alphaa)^2\label{2cho}.\end{equation} The eigenvalues of
(\ref{2cho}) coincide with the eigenvalues associated with
antisymmetric eigenfunctions of the double oscillator Hamiltonian,
\begin{equation}\label{robson}-{\d^2\over\d x^2}+(|x|-\aa)^2.\end{equation}
In \cite[Table  1]{robson}, the two lowest eigenvalues of this
operator are tabulated. We are interested in the second one since it
corresponds to the groundstate of $h^2_{KG0}$.

Of course, quarks are coupled not only with the scalar field, but
also with the omega field. This coupling was not included in eq.
(\ref{h_D}) because, as explained in \cite{guichon,guichon2}, its
only effect is  the nucleon-omega coupling which will be considered
in due course.


The simplified model based on eq. (\ref{BKG}) is  interesting on its
own right because the essential physics is already contained in it.
Moreover, it seems likely that Bogoliubov's model originated from
the consideration of the corresponding Klein-Gordon equation.
The exact groundstate eigenvalue of $h^2_{KG0}$ (eq. (\ref{2cho}))
may be easily determined. For $0\leq\alphaa\leq2,$ the relevant
values in \cite[Table  1]{robson} are reproduced with the desired
accuracy by the following interpolating expression,
\begin{eqnarray}&&
{m^2(\kappa,\aa)\over\kappa} =3. - 2.2553\aa + 0.77444\aa^2 -
0.032099\aa^3 - 0.018519\aa^4. \label{m^2-alpha-00}\end{eqnarray}
The presence of an external scalar field $\sigma$ is described
through the replacement $\alphaa= g^q_s\sigma/\sqrt{\kappa}.$
The groundstate eigenvalue of $h_{KG}^2$ is denoted by
$m^2(\kappa,g^q_\sigma\sigma/\sqrt{\kappa})$.


\section{Ansatz}
For the groundstate of $h^2_{KG0}$, eq. (\ref{2cho}), we also
consider the ansatz
\begin{equation}\Psi_{\betab}=\exp\left(-{1\over2}(|{\bf
r}|-\alphaa-\betab)^2\right),\label{ansatz}\end{equation}where the
parameter $\betab$ should be fixed variationally for each value of
$\alphaa$. With $\betab=0$, this ansatz had already been proposed in
\cite{bohr} and is exact for $a=0$ and in the limit
$\alphaa\rightarrow\infty.$
\subsection{Simplified model}
The ansatz $\Psi_\betab$ provides a very good approximation to the
groundstate eigenvalue of $h^2_{KG0}$. 
We have,
\begin{equation}\label{m^2-alpha-S}{m^2(\kappa,\aa)\over\kappa}=
\min_\betab{\langle\Psi_{\beta}|h^2_{KG0}|\Psi_{\betab}\rangle\over\langle\Psi_{\betab}|\Psi_{\betab}\rangle}
=\min_\betab{{\cal K}_0+{\cal V}_0\over {\cal
N}_0},\end{equation}where
\begin{eqnarray*}&&{\cal N}_0=
   \pi\left(2(\aa + \bb)\e^{-(\aa + \bb)^2} + \left(1 + 2(\aa+\bb)^2\right)
\sqrt\pi\left(1+\Erf(\aa + \bb)\right)\right)\\
&&{\cal V}_0=  { 1\over2} \pi\left[2 (\aa +5\bb + 2\aa\bb^2 +
2\bb^3)\e^{-(\aa + \bb)^2}\right.\\&&\quad\left. +\left(3 + 12\bb^2
+ 4\bb^4 + 4\aa\bb(3 + 2\bb^2) + \aa^2(2 + 4\bb^2)\right)
 \sqrt\pi(1+\Erf(\aa + \bb))\right]\\
&&
{\cal K}_0=  {1\over2} \pi \left[2(\aa + \bb)\e^{-(\aa + \bb)^2} +(3
+ 2\aa^2 +4\aa\bb + 2\bb^2)
 \sqrt \pi(1+\Erf(\aa + \bb))\right].\end{eqnarray*}
Minimization of the r.h.s. of eq. (\ref{m^2-alpha-S}) with respect
to $\bb$ is easily performed.  {The remarkable agreement between the
exact eq. (\ref{m^2-alpha-00}) and the variational eq.
(\ref{m^2-alpha-S}) deserves to be  stressed.}
\begin{table*} 
\caption{Numerical results  for $m^2(\kappa,\alphaa)/\kappa$ taking
  distinct values of $a$. Expression
(\ref{m^2-alpha-S}), based on eq. (\ref{ansatz}), is compared with the
corresponding exact results obtained from eq. (\ref{m^2-alpha-00}),
The minimizing value of $\betab$ is shown. {The agreement between
the exact eq. (\ref{m^2-alpha-00}) and the variational eq.
(\ref{m^2-alpha-S}) is very good.} \vspace*{0.5cm}}\label{tab1}
\centering
\begin{tabular}{l|c|c|c}\hline\hline
$\alphaa$ &$m^2(\kappa,\alphaa)/\kappa$&$m^2(\kappa,\alphaa)/\kappa$&$\betab$\\
&eq. (\ref{m^2-alpha-00})&eq.
(\ref{m^2-alpha-S}).& \\\hline
0&3&3&0\\0.1&2.7821&2.7822&-0.05368 \\0.2&2.5796&2.5801&  -0.1053\\0.3&2.3921&2.3932&-0.1547 \\0.4&2.2193&2.2213& -0.2017\\
0.5&2.0608& 2.0641&-0.246\\0.6&1.9163& 1.9210&-0.2875
\\0.7&1.7853&1.7917
&-0.326\\0.8&1.6674&1.6756&-0.3612\\0.9&1.5620&1.5723& -0.39283\\
1&1.4685&1.481&-0.42068\\1.1&1.3864&1.4011& -0.44453
\\1.2&1.3150&1.3319& -0.46418
\\\hline
\end{tabular}
\end{table*}
\subsection{Full model}
In order to estimate the lowest eigenvalue of 
\begin{equation}\label{2D0} h_{D0}^2=-\nabla^2+(|\bfr|-
a)^2+i\beta\bfalpha\cdot{\bfr\over|\bfr|},\end{equation} the Dirac
structure of the quark wave-function, which has been neglected in
the previous ansatz $\Psi_\betab$, should be taken into account, so
that we consider a new ansatz,
\begin{equation}\label{newansatz}\Psi_{\betab,\lambda}=\left[\begin{matrix}\chi
\\i\lambda(\bfsigma\cdot\bfr)\chi\end{matrix}\right]\e^{-(|r|-\aa-\bb)^2/2},\end{equation}
where ${\betab,\lambda}$ are variational parameters, and $\chi$ is a
2-spinor. Minimizing the expectation value of $h_{D0}^2$ for
$\Psi_{\betab,\lambda},$ the following expression for the quark mass
is found,
\begin{equation}{m^2(\kappa,\aa)\over\kappa}=\min_{\lambda,\bb}{\langle \psi_{\betab,\lambda}|h_{D0}^2|\psi_{\betab,\lambda}\rangle
\over\kappa\langle
\psi_{\betab,\lambda}|\psi_{\betab,\lambda}\rangle}=
\min_{\lambda,\bb}{{\cal K}_0+{\cal V}_0+{\cal V}_{01}\la+({\cal
K}_1+{\cal V}_1)\la^2\over {\cal N}_0+{\cal
N}_1\la^2},\label{m^2/K}\end{equation} where ${\cal N}_0$, ${\cal
V}_0$ and ${\cal K}_0$, are as in the previous subsection and
$${\cal N}_1={1\over2} \pi \left(
\left(2(\aa+\bb)(5  + 2(\aa+\bb)^2\right) \e^{-(\aa+\bb)^2} + (3 +
12(\aa+\bb)^2 + 4(\aa+\bb)^4) \sqrt\pi(1+\Erf(\aa+\bb))\right)$$
$${\cal V}_1={1\over4} \pi\left\{\left[2(33\bb +28\bb^3 + 4\bb^5 + \aa^3(2 + 4\bb^2) +
2\aa^2\bb(11 + 6\bb^2) + \a(17 + 48\bb^2 + 12\bb^4))\right]
\e^{-((\aa + \bb))^2}\right.$$
$$ + \left[(15 + 90\bb^2 +
60\bb^4 + 8 \bb^6 + 16\aa^3\bb(3 + 2\bb^2) + \aa^4(4 + 8\bb^2) +
12\aa^2(3 + 12\bb^2 + 4\bb^4)\right.$$
$$\left.\left. + 8\a\b(15 +
20\bb^2 + 4\bb^4)) \sqrt \pi\right](1+\Erf(\aa + \bb))\right\}$$
$${\cal K}_1={1\over4} \pi \left(
+ \left((34(\aa+\bb) +4(\aa+\bb)^3)\e^{-(\aa+\bb)^2}
 + (15 + 36(\aa+\bb)^2 +
4(\aa+\bb)^4) \sqrt\pi\right)(1+\Erf(\aa+\bb))\right)$$
$${\cal V}_{01}=
 -2\pi \left(2(1 + (\aa+\bb)^2)\e^{-(\aa+\bb)^2} +
(\aa+\bb)(3 + 2(\aa+\bb)^2) \sqrt\pi(1+\Erf(\aa+\bb))\right).$$
Minimization of eq. (\ref{m^2/K}) w.r.t. $\lambda$ is readily
performed, so that
\begin{equation}{m^2(\kappa,\aa)\over\kappa}={1\over2}\min_\betab\left({{\cal K}_0+{\cal V}_0\over{\cal N}_0}+{{\cal K}_1+{\cal V}_1\over{\cal N}_1}
-\sqrt{\left({{\cal K}_0+{\cal V}_0\over{\cal N}_0}-{{\cal
K}_1+{\cal V}_1\over{\cal N}_1}\right)^2+\left({{\cal
V}_{01}\over\sqrt{{\cal N}_0{\cal
N}_1}}\right)^2}~\right)\label{m^2-alpha}.\end{equation}
Minimization of the r.h.s. of eq. (\ref{m^2-alpha}) with respect to
$\bb$ may be easily implemented. In order to obtain the EoS, we need
the quark mass under the effect of an external scalar field,
$m(\kappa,g_\sigma\sigma/\sqrt\kappa)$. We observe that the
simplified model considered in the previous section is recovered by
setting $\lambda=0$ in eq. (\ref{m^2/K}), instead of performimg the
indicated minimization with respect to $\lambda$.


\begin{figure}[ht]
\vspace{1.5cm} \centering
\includegraphics[width=0.8\linewidth,angle=0]{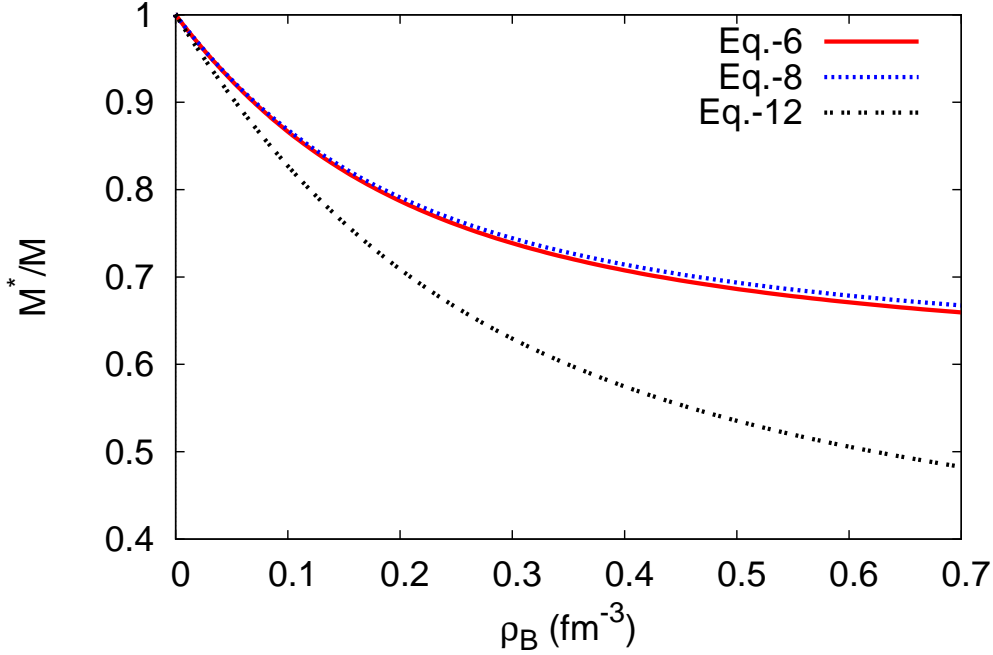}
\caption{
$M^*/M$ for the {present} QMC approach based on the Bogolyubov
model. Improved calculation according to eqs. (\ref{m^2-alpha-00}),
(\ref{m^2-alpha-S}) and (\ref{m^2-alpha}).
 } \label{Fig2}
\end{figure}
\section{QMC model. Bogoliubov model with external scalar field} According to the
QMC model \cite{guichon}, nuclear matter is a system of nucleons
which behave like point-like particles, although they are
constituted by quarks coupled to the scalar $\sigma$ field, in the
framework of an independent particle model. The QMC model, was
proposed by Guichon \cite{guichon} on the basis of the MIT bag model
\cite{chodos} and has been considered by other authors
\cite{guichon1,whittenbury,batista,barik1}.
Here, we wish to implement the QMC model based on the Bogoliubov
quark model \cite{bogolubov}, which describes the quark dynamics in
the nucleon rest frame.

\begin{figure}[ht]
\vspace{1.5cm} \centering
\includegraphics[width=0.8\linewidth,angle=0]{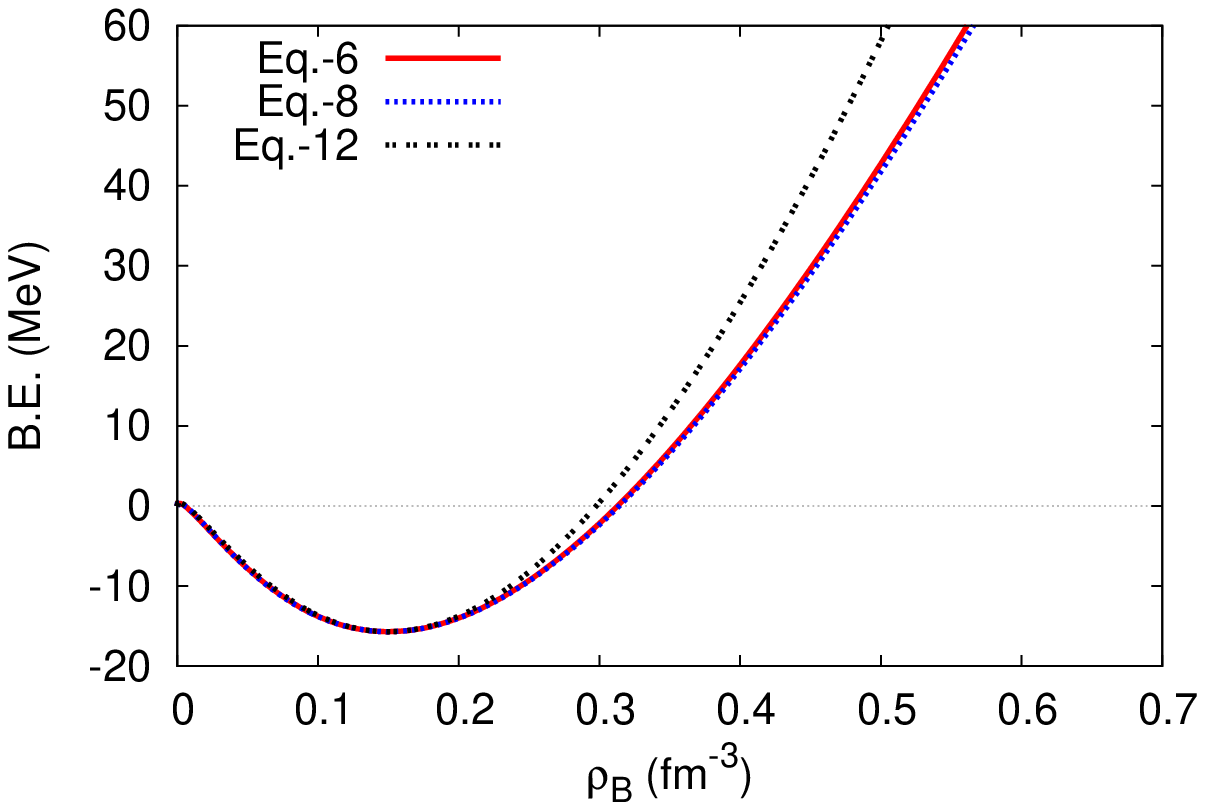}
\caption{ EoS for the {present} QMC approach based on the Bogolyubov
model. Improved calculation according to eqs. (\ref{m^2-alpha-00}),
(\ref{m^2-alpha-S}) and (\ref{m^2-alpha}), with $\betab=0$ and with
optimal $\betab$. It is important to minimize the rhs. of eq.
(\ref{m^2-alpha}) w.r.t. $\betab$.
 } \label{Fig3}
\end{figure}

The energy density of quark matter reads
\begin{eqnarray}&&{\cal E}={\gamma\over(2\pi)^3}\int^{k_F}\d^3k\sqrt{k^2+{M^*}^2}+{1\over2}m_\sigma^2\sigma^2
-{1\over2}m_\omega^2\omega^2+3g^q_\omega\omega\rho_B\label{eq3}\end{eqnarray}
where {$g^q_\omega$ denotes the quark-$\omega$ coupling},
\begin{eqnarray}&&{\rho_B}={\gamma\over(2\pi)^3}\int^{k_F}\d^3k,\quad
M^*=3{m}_q(\sigma)=3m(\kappa,{g_\sigma^q\sigma/\sqrt{\kappa}}
)\nonumber\end{eqnarray} and $\gamma=4$ denotes the spin isospin
degeneracy.
The pressure is given by
$${-P}={\gamma\over(2\pi)^3}\int^{k_F}\d^3k\sqrt{k^2+{M^*}^2}+\rho_B
\left(3g^q_\omega\omega-\sqrt{k_F^2+{M^*}^2}\right)+{1\over2}m_\sigma^2\sigma^2
-{1\over2}m_\omega^2\omega^2.
$$

Minimization of $\cal E$ with respect to $\sigma$ and $\omega$ is expressed by
the conditions
\begin{equation}
{\partial{\cal E}\over\partial\sigma}={\partial
M^*\over\partial\sigma}\rho_s+ m^2_\sigma\sigma =0,
\label{sigma}
\end{equation}
where$$ \rho_s=
{\gamma\over(2\pi)^3}
\int^{k_F}\d^3k{M^*\over\sqrt{k^2+{M^*}^2}},$$
denotes the scalar
density, and
$$m_\omega^2\omega =3g_\omega^q\rho_B.$$
Since the factor $ {\partial
M^*/\partial\sigma}:=-3g^q_s{\cal S}_G$ is density dependent,
equation (\ref{sigma}) shows that the source of the $\sigma$ field is not simply
the scalar density, as already remarked in \cite{guichon,guichon2}.
The nucleon-$\sigma$ coupling, which is described by  $-{\partial
M^*/\partial\sigma}$, decreases with density. Indeed, ${\cal S}_G$
decreases fast as density increases, being ${\cal S}_G=1$ at the
vacuum.

\begin{table*} 
\caption{Numerical results for $m^2(\kappa,\alphaa)/\kappa$,
corresponding to distinct values of $\alphaa$ comparing the
simplified model, according to eq. (\ref{m^2-alpha-00}) and
(\ref{m^2-alpha-S}), with the full model given by equation (\ref{m^2-alpha}).
\vspace*{0.5cm}}\label{tab2} \centering
\begin{tabular}{l|c|c|c}\hline\hline
$\alphaa$& $m^2(\kappa,\alphaa)/\kappa$&$m^2(\kappa,\alphaa)/\kappa$&$m^2(\kappa,\alphaa)/\kappa$\\
&eq. (\ref{m^2-alpha-00})&eq. (\ref{m^2-alpha-S}).&eq.
(\ref{m^2-alpha})\\\hline
0&3&3&2.64022\\0.1&2.7821&2.78233&2.41146\\0.2&2.5796& 2.58006& 2.19766\\0.3&2.3921&2.39318,& 1.99851\\0.4&2.2193&2.22133& 1.81364\\
0.5&2.0608&2.06408&1.64265\\0.6&1.9163&1.92101&
1.48508\\0.7&1.7853&1.7917
& 1.34047\\0.8&1.6674&1.67564& 1.20829\\0.9&1.5620&1.57228& 1.08799\\
1&1.4685&1.481&0.978984\\1.1&1.3864&1.40112&
0.880623\\1.2&1.3150&1.33192& 0.79224
\\1.3&1.2153& 1.2726&0.713139\\1.4&1.2013& 1.2223&0.642597\\1.5&1.1575& 1.18013&0.579884\\1.6&1.1213&
1.14522&0.52427
\\1.7&1.0919&  1.11679&0.475039\\1.8&1.0684&
1.0941&0.431503\\1.9&1.0500&1.07602&0.393011
\\2&1.0358&1.0599&0.358953
\\\hline
\end{tabular}
\end{table*}

\color{black} We observe that
eq. (\ref{eq3}) for the binding energy is based on the so-called
mean field approximation. This means that it contains only one body
terms, and that the two-body interactions are simulated by the
quadratic terms in $\sigma^2$ and $\omega^2$, namely, $ m_\sigma^2
\sigma^2/2$ and $- m_\omega^2 \omega^2/2$. Notice also that $M^*$
depends on $\sigma$, but not on $\omega$.
The
quark mass is obtained from
$m_q(\sigma)=M^*/3=m(\kappa,g_\sigma\sigma/\sqrt{\kappa})$.


It may be observed that, in the QMC model proposed by Guichon, the
sigma-quark coupling takes values that go from less than $6$ to
about $8$, see \cite{guichon1,panda}. The obtained value in the
present model is only slightly smaller. The difference, which is not
dramatic, may be due to using a string-type confining potential
instead of the MIT bag model.

\color{black}

For $0<\alphaa<2$, it may be safely assumed that$$M^* =\sqrt{ 9
(2.6402 - 2.3634 \ala + 0.7631 \ala^2 - 0.045119 \ala^3 - 0.015778
\ala^4) \kappa}.$$ This expression fits the minimization with
respect to $b$ indicated in the r.h.s. of eq.(\ref{m^2-alpha}).
\color{black} Thus,
$$3g^q_s{\cal S}_G=-{\partial M^*\over\partial\sigma}=-{\partial
M^*\over\partial\ala}~{g^q_s\over\sqrt{\kappa}}.$$\color{black}

\begin{table*} 
\caption{Model parameters and nuclear matter properties at saturation
  taking the binding
energy 15.7 MeV, the equilibrium density 0.15 fm$^{-3}$ and the
vacuum quark mass 313 MeV, and  effective mass $M_N^*$ calcuted using
Eqs. (\ref{m^2-alpha-00}), (\ref{m^2-alpha-S}) and  (\ref{m^2-alpha})
\vspace*{0.5cm}}
\label{tab3} \centering
\begin{tabular}{l|c|c|c}\hline\hline
&eq. (\ref{m^2-alpha-00})&eq. (\ref{m^2-alpha-S}).&eq. (\ref{m^2-alpha})\\\hline
$g_\sigma^q$&4.0562&4.0339&3.9851
\\{$g_\omega=3g_\omega^q$}&7.6533&7.5530&9.3085
\\$\kappa$ (MeV$^2$)&32656.&32656.&37106.
\\$\sigma$ (MeV)&23.357&23.113&27.490
\\$\omega$ (MeV)&14.387&14.199&17.499
\\$M_N^*$ (MeV)&771.36&774.39&715.51
\\$K$ (MeV) &293.64&297.42&335.17
\\\hline
\end{tabular}
\end{table*}
By including $\rho$ mesons, the present formalism may easily be
extended to the treatment of asymmetric nuclear matter,
\begin{eqnarray}&&{\cal E}={\gamma\over(2\pi)^3}\sum_{j=P,N}\int^{k^j_F}\d^3k\sqrt{k^2+{M^*}^2}+{1\over2}m_\sigma^2\sigma^2
-{1\over2}m_\omega^2\omega^2+3g^q_\omega\omega\rho_B-{1\over2}m_b^2b^2+g^q_bb\rho_3\nonumber\\\label{asym}\end{eqnarray}
where {$g^q_b$ denotes the quark-$\rho$ coupling} and
\begin{eqnarray}&&{\rho_B}={\gamma\over(2\pi)^3}\sum_{j=P,N}\int^{k^j_F}\d^3k,\quad
{\rho_3}={\gamma\over(2\pi)^3}\sum_{j=P,N}\eta_j\int^{k^j_F}\d^3k,\quad
M^*=3{m}_q(\sigma)=3m(\kappa,{g_\sigma^q\sigma/\sqrt{\kappa}}
)\nonumber.\end{eqnarray}Here, $k^j_F,~j=P,N$ denote the Fermi
momenta of protons and neutrons,  and $\gamma=2$ denotes the spin
degeneracy.
\begin{figure}[ht]
\vspace{1.5cm} \centering
\includegraphics[width=0.8\linewidth,angle=0]{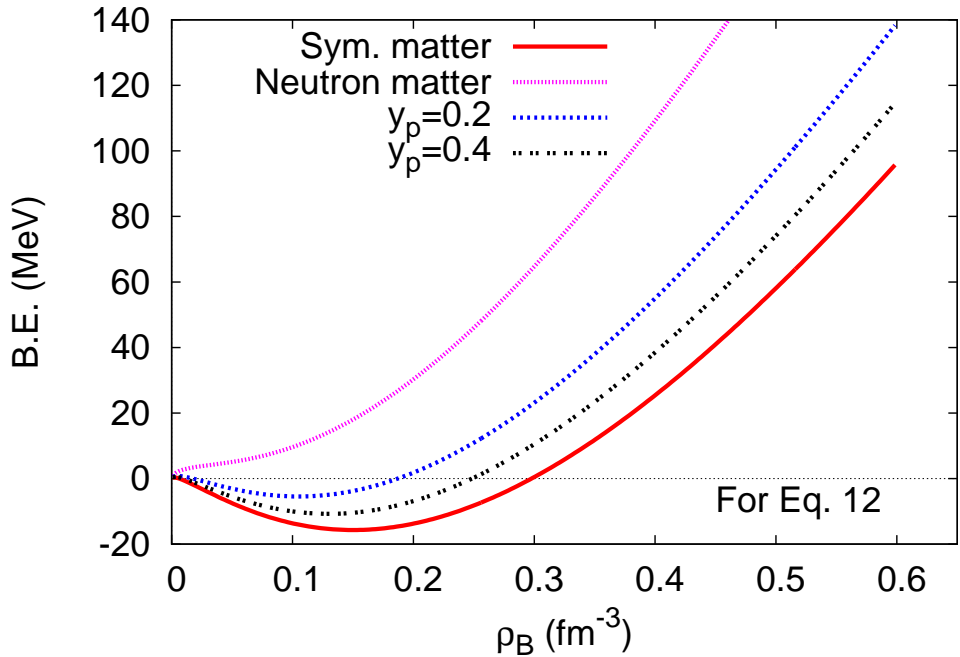}
\caption{ EoS of asymmetric nuclear matter for several proton
fractions, according to the {present} QMC approach based on the
Bogolyubov model.  Calculation according to
eq. (\ref{m^2-alpha}).
} \label{Fig6}
\end{figure}

\section{Discussion}

In the present work we have proposed an effective relativistic
nuclear model that takes explicitly into account the internal
structure of the nucleon, in the spirit of the QMC model proposed by
Guichon \cite{guichon}. Hadronic matter at normal densities is a
system of nucleons composed of quarks coupled to $\sigma$ and
$\omega$ fields and bound by a linearly raising potential  according
to the Bogoliubov model of baryons \cite{bogolubov} and  as
suggested by gauge theories. The parameters of the model have been
fitted to the saturation density and the binding energy of symmetric
nuclear matter, and to the nucleon mass in vacuum.

In Table \ref{tab1} we compare the exact groundstate eigenvalue of
$h^2_{KG0}$ which determines the effective mass of the nucleon, eq. (\ref{2cho}), with the variational result based on
the ansatz (\ref{ansatz}), showing the value of the parameter
$b$ which optimizes the variational result. The performance of
the ansatz is remarkable. In Table \ref{tab2}, the exact groundstate
energy of $h^2_{KG0}$, the corresponding variational approximation
for ansatz (\ref{ansatz}) and the variational groundstate energy of
$h^2_{D0}$ for the new ansatz (\ref{newansatz}) are tabulated for
diffarent values of $a$. We remark that the expectation value
of $h^2_{D0}$ for the ansatz (\ref{newansatz}) with $\lambda=0$
coincides with the expectation value of $h^2_{KG0}$ for the ansatz
(\ref{ansatz}). We conclude that the term of $h^2_{D0}$ involving
the Dirac matrix $\beta$ makes a significant difference and that the
Dirac structure of the quark wave function should not be neglected,
although the model based on $h^2_{KG0}$ performs satisfactorily and
is physically meaningful.

\begin{table*} 
\caption{Neutron star properties calculated with the  EOS obtained
  from Eqs.
  (\ref{m^2-alpha-00}), (\ref{m^2-alpha-S}) and (\ref{m^2-alpha}):maximum gravitational mass
$M_{max}$, maximum baryonic mass $M_{b~max}$ and radius $R$.
\vspace*{0.5cm}}\label{tab4} \centering
\begin{tabular}{l|c|c|c}\hline\hline
&eq. (\ref{m^2-alpha-00})&eq.
(\ref{m^2-alpha-S}).&eq. (\ref{m^2-alpha})\\\hline $M_{max}$
[$M_{\odot}$]& 2.22&2.21&2.40
\\$M_{b~max}$ [$M_{\odot}$]&2.59& 2.58&2.83
\\$R$ [Km]&12.25&  12.21&12.65
\\\hline
\end{tabular}
\end{table*}

The curves displayed in Figure \ref{Fig3}, representing the binding
energy vs. the baryon density, were obtained using eqs.
(\ref{m^2-alpha-00}), (\ref{m^2-alpha-S}) and (\ref{m^2-alpha}). The
model parameters are $m_\sigma=$550 MeV, $m_\omega=783$ MeV and the
quark mass for nucleons in vacuum, $m_q=313$ MeV. The coupling
constants $g_\sigma^q,~g_\omega^q$ were chosen so as to reproduce
the binding energy and density at equilibrium, that is, ${\cal
E}/\rho_B - M_N = -15.7$~MeV at saturation
$\rho_B=\rho_0=0.15$~fm$^{-3}$ (pressure $P=0$), being $M_N=939$ MeV
the free nucleon mass. The value of $\kappa$ is determined by the
quark mass  in vacuum. We find that eqs. (\ref{m^2-alpha-00}) and
(\ref{m^2-alpha-S}) are associated with $\kappa=32656.3$ MeV$^2$,
while eq. (\ref{m^2-alpha}) is associated with $\kappa=3 7106.6$
MeV$^2$. Taking the Dirac structure of the quark wave function into
account, leads to an increase of the string tension. At saturation
density, the effective mass is given by $M^*/M=0.821$ for
eq.(\ref{m^2-alpha-00}), $M^*/M=0.825$ for eq.(\ref{m^2-alpha-S})
and $M^*/M=0.762$ and for eq.(\ref{m^2-alpha}), while the
incompressibility is given by $K=293.64$ MeV for  eqs.
(\ref{m^2-alpha-00}), by  $K=297.42$ for (\ref{m^2-alpha-S}) and by
$K=335.17$ MeV for eq. (\ref{m^2-alpha}). We point out that for the
QMC model based on the MIT bag model, the obtained incompressibility
is 291 MeV.

The outputs are summarized in Table \ref{tab3}. We
observe that the values we find for the incompressibility are quite
close to what might be called the ``empirical value" obtained by
Satpathy and Nayak \cite{satpathy}, $K = 288\pm 20$ MeV, however, too
high for the value obtained in \cite{vidana}  where the value $K =
230\pm 40$ MeV constrained by  the giant monopole resonance, is obtained. It appears
that the more correct treatment of the Dirac model gives too large
an incompressibility. However, it is likely that, by using a more
flexible variational ansatz, for instance, one with different $b$'s
in the upper and in the lower components of the Dirac wavefunction,
the stiffness will be slightly reduced. \color{black}

\begin{figure}[ht]
\vspace{1.5cm} \centering
\includegraphics[width=0.8\linewidth,angle=0]{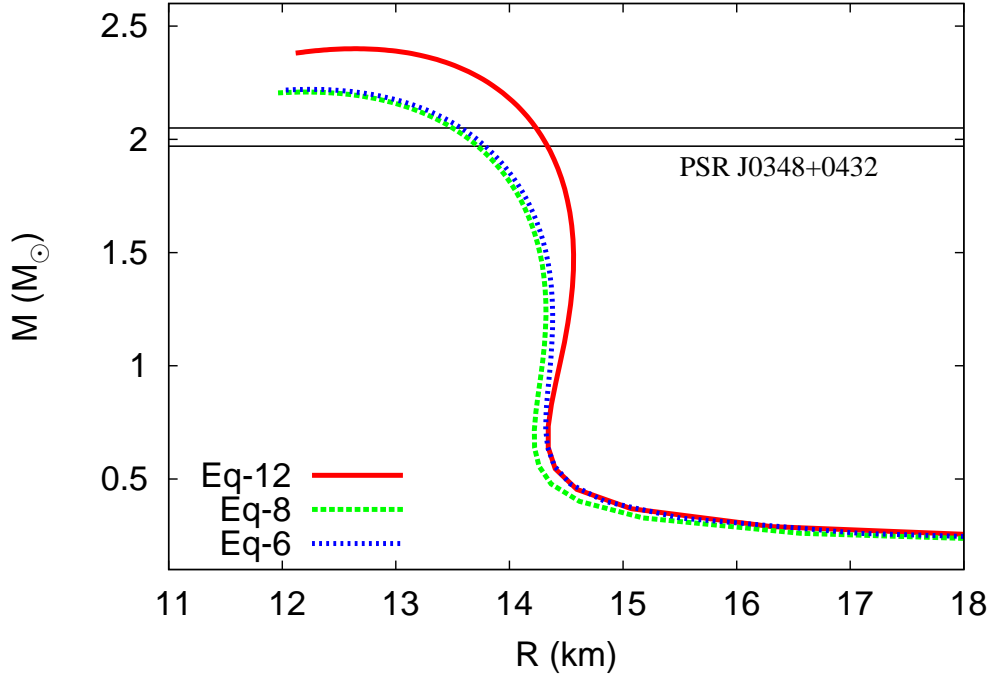}
\caption{Gravitational mass vs. radius  for a neutron star. The
constraint imposed by the pulsar PSR J0348+0432 with a mass
$2.01\pm0.04\, M_\odot$ \cite{antoniadis_13} is represented.
Different curves were obtained calculating the  effective mass of
the nucleon from Eqs. (\ref{m^2-alpha-00}), (\ref{m^2-alpha-S}) and
(\ref{m^2-alpha}).
} \label{Fig7}
\end{figure}

Asymmetric nuclear matter is described by including $\rho$ meson,
the $\rho$-quark coupling $g_b^q$ having been chosen so as to
reproduce a symmetry energy of $32$ MeV at saturation. In Figure
\ref{Fig6}, the EoS for asymmetric matter is represented for several
proton fractions calculated with   (\ref{m^2-alpha}). As expected,
the binding energy increases with an increase of isospin asymmetry
and neutron matter is not bound.

Next we have considered the EoS of stellar matter including
electrons and muons and imposing $\beta$-equilibrium and charge
neutrality in order to calculate the structure of neutron stars
within the present model. Two solar mass pulsars have been recently
confirmed, see \cite{demorest_10,antoniadis_13}, and acceptable
models should describe stars with a mass  at least as large as
$2M_\odot$.

We have integrated the  Tolmann-Oppenheimer-Volkoff equations of
hydrostatic equilibrium in general relativity \cite{tov} and
computed neutron star properties, namely, the maximum gravitational
mass $M_{max}$, the maximum baryonic mass $M_{b~max}$ and the
respective radius $R$, shown in Table \ref{tab4} for the different
approximations discussed in the previous sections. For the outer
crust the Baym-Pethick-Sutherland EoS (BPS) \cite{Baym} was used. In
Figure \ref{Fig7}, we plot  the mass versus the radius of the family
of stars described with the stellar matter equation of state
obtained for the nucleon effective mass calculated from  Eqs.
(\ref{m^2-alpha-00}), (\ref{m^2-alpha-S}) and (\ref{m^2-alpha}).
Since the full model with the mass given by (\ref{m^2-alpha}) gives
rise to the hardest EoS, it is not surprising that the full model
predicts the largest masses. In  Figure \ref{Fig7} we also include
the mass constraints imposed by the  the pulsar PSR J0348+0432 with
a mass  $2.01\pm0.04\, M_\odot$ \cite{antoniadis_13}.


The following considerations are in order. The tendencies to chiral
symmetry restoration and deconfinement show up in the increasing
nucleon bag size, and in the decreasing effective mass, for
increasing density. Indeed, eq. (\ref{ansatz}) suggests that the
nucleon bag looks like a hollow bubble, whose radius $a+b$,
($a=g_\sigma^q\sigma/\sqrt\kappa$), increases with density, becoming
rather large at high densities. Notice that $0\leq a+b\leq a$. At
extremely low and extremely high densities, $b$ becomes equal to
$0$. On the other hand, according to Fig. \ref{Fig2}, the mass
decreases appreciably as the density increases.

The  obtained phenomenological string tension is somewhat low. This
is because we have considered the oversimplification of an
independent particle model, i.e., a one-body force, to describe the
three body force which  represents  the actual string interaction
connecting each quark in the nucleon to the common center of mass. A
reduction in the phenomenological string tension is the price payed
for the extreme simplicity of the used model. Notice that strings
are one dimensional objects, and we use a three dimensional
confining one-body potential to represent them.
\section{Conclusions}
In summary, we have proposed a relativistic nuclear model that takes
the quarks as fundamental constituents, and describes the nucleons
as composite particles. The quarks interact in the vacuum through a
linear interaction, and medium effects are taken into account
through the coupling of the quarks to mesons fields. The mesonic
fields are obtained through a minimization of the thermodynamical
potential. The parameters of the model are chosen so that saturation
nuclear matter properties are described, however, the
incompressibility comes a bit too high. As soon the internuclear
spacing is of the order of the radius of the nucleon a phase
transition to quark matter is expected.
An interesting and  consistent EoS embodying the hadron-quark phase
transition may be obtained by describing the quark phase in the
framework of Bogoliubov's independent quark model. \color{black}
This will be considered in the near future. The effect of the
temperature in the behavior of the radius will also be investigated.

\section*{Apendix }
Minimization of the r.h.s. of eq. (\ref{m^2-alpha-S}) with respect
to $\bb$ may be 
taken into account with sufficient accuracy by the replacement, for
$0\leq\alphaa<2,$
$$\beta=c_1\aa+c_2\aa^2+c_3\aa^3+c_4\aa^4+c_5\aa^5+c_6\aa^6,$$where
$c_1 = -0.54690, c_2 = 0.096421, c_3 = 0.024973, c_4 =-0.00001286,
c_5 = 0.008702,$ $ c_6 = -0.003858$. It is also well described by
assuming that, for
$0\leq\alphaa<2,$ 
$${m^2(\kappa,\alpha)\over\kappa }=3. -2.256\aa+ 0.78923\aa^2 -0.03179\aa^3 -0.020473\aa^4.
$$

Minimization of the r.h.s. of eq. (\ref{m^2-alpha}) with respect to
$\bb$ may be 
taken into account with sufficient accuracy by the replacement,  for
$0\leq\alphaa<2,$
$$\beta=d_0+d_1\aa+d_2\aa^2+d_3\aa^3+d_4\aa^4+d_5\aa^5+d_6\aa^6,$$ where
$d_0 = 0.00805, d_1 = -0.53949, d_2 = 0.093915, d_3 = 0.024846, d_4
= 0.0029149, d_5 = 0.00068587, d_6 = -0.0013336.$
It is also well described by assuming that,  for $0\leq\alphaa<2,$
$${m^2(\kappa,\alphaa)\over\kappa}=2.6402 -2.3634\aa +0.7631\aa^2 -0.045119\aa^3 -0.015778\aa^4
.$$

 \color{black}


\end{document}